\documentstyle[epsf,12pt]{article}

\begin{document}
\begin{center}

{\Large\bf Covariant tensor formalism for partial wave analyses of
$\psi$ decays into $\gamma B\bar B$}
 \vskip 0.5cm

{\large\bf Sayipjamal Dulat$^{b,c}$, Bo-Chao Liu$^{b}$ ,
 Bing-Song Zou$^{a,b,}$, and
Ji-Min Wu$^{a,b,}$}

a) CCAST (World Laboratory), P.O.~Box 8730, Beijing 100080\\
b) Institute of High Energy Physics, CAS, P.O.~Box 918(4),
Beijing, 100039,
China\\
c)Department of Physics, Xinjiang University, Urumqi, 830046,   China \\
\vskip 0.5cm
\end{center}

\begin{abstract}
Recently BES2 collaboration observed an enhancement near the
$p\bar p$ invariant  mass spectrum. Using the covariant tensor
formalism, here we provide theoretical formulae for the partial
wave analysis (PWA) of the $\psi$ radiative decay channels $\psi
\to \gamma p\bar p, \gamma\Lambda\bar \Lambda, \gamma\Sigma
\bar\Sigma, \gamma\Xi\bar \Xi$ . By performing the Monte Carlo
simulation, we give the angular distributions for the photon and
proton in the process of $J/\psi \to \gamma p\bar p$, which may
serve as a useful reference for the future PWA on these channels.
\end{abstract}

\section{Introduction}
$J/\psi$ and $\psi'$ radiative decay to $B\bar B$ (baryon and
antibaryon pair) is a good channel to study the possible bound or
resonant states of the $B\bar B$ system. Abundant $J/\psi$ and
$\psi'$ events have been collected at the Beijing Electron
Positron Collider (BEPC). More data will be collected at upgraded
BEPC and CLEO-C.  Based on the analysis of the 58 million $J/\psi$
events accumulated by the BES2 detector at the BEPC, recently BES2
  reported \cite{BES} that they observed
a strong, narrow enhancement near the threshold in the invariant
mass spectrum of $p\bar p$ (proton - antiproton) pairs from
$J/\psi\to \gamma p\bar p$ radiative decays.  The structure has
the properties consistent with either an $S(0^{-+})$- or
$P(0^{++})$-wave Breit-Wigner resonance function. In the S-wave
case, the peak mass is blow $2M_p$ around $M= 1859MeV$ with a
total width $\Gamma<30MeV$ .In order to get more useful
information about properties of the resonances such as their
$J^{PC}$ quantum numbers, mass, width, production and decay rates,
etc., partial wave analyses (PWA) are necessary. PWA is an
effective method for analysing the experimental data of hadron
spectrum. There are two types of PWA: one is based on the
covariant tensor (also named Rarita-Schwinger)
formalism\cite{rarita} and the other is based on the helicity
formalism\cite{Chung}. Ref.\cite{filip} showed the connection
between the covariant tensor formalism and the covariant helicity
one. Ref.\cite{zou} provided PWA formulae in a covariant tensor
formalism for $\psi$ decays to mesons, which have been used for a
number of channels already published by
BES\cite{BES1,BES2,BES3,BES4,BES5,BES6} and are going to be used
for more channels. A similar approach has been used in analyzing
other reactions\cite{Anisovich1,Anisovich2,Anisovich3}.
Ref.\cite{zou2} provided explicit formulae for the angular
distribution of photon of the $\psi$ radiative decays in the
covariant tensor formalism, and also discussed  helicity formalism
of the angular distribution of the $\psi$ radiative decays to two
pseudoscalar mesons, and its relation to the covariant tensor
formalism. Now we extend the covariant tensor formalism to the
$\psi \rightarrow \gamma B\bar B$ with B represents baryons  $p$,
$\Lambda$, $\Sigma$, $\Xi$ etc, and derive the decay amplitudes
for various intermediate resonant states in the framework of the
relativistic covariant tensor formalism.

In this paper we study the phenomenological spin parity
determination of resonances. The plan of this article is as
follows: in section 2, we present the necessary tools for the
calculation of the tensor amplitudes, within a covariant tensor
formalism. This will allow us to derive covariant amplitudes for
all possible processes.  In section 3, we present covariant tensor
formalism for $\psi$ radiative decays to baryon antibaryon pairs.
Since covariance is a general requirement of any decay amplitude,
all possible amplitudes  are written in terms of covariant tensor
form. All amplitudes include a complex coupling constant and
Blatt-Weisskopf centrifugal barriers where necessary. In section
3, we provide the angular distribution of the photon and proton.
The conclusions are given in section 5.

\section{Prescriptions for the construction of covariant tensor
amplitudes}

 In this section we present the necessary tools for the
construction of covariant tensor amplitudes. The partial wave
amplitudes $U_i^{\mu\nu\alpha}$ in the covariant Rarita-Schwinger
tensor formalism  can be constructed by using pure orbital angular
momentum covariant tensors $\tilde
t^{(L_{bc})}_{\mu_1\cdots\mu_{L_{bc}}}$ and covariant spin wave
functions $\phi_{\mu_1 \cdots\mu_s}$ together with metric tensor
$g^{\mu\nu}$,  totally antisymmetric Levi-Civita tensor
$\epsilon_{\mu\nu\lambda\sigma}$ and momenta of parent particles.
 For a process $a\to bc$, if there exists a relative orbital
angualar momentum $\textbf{L}_{bc}$ between the particle a and b, then the
 pure orbital angular momentum
 $\textbf{L}_{bc}$ state can be represented by
covariant tensor wave functions $\tilde
t^{(L_{bc})}_{\mu_1\cdots\mu_{L_{bc}}}$\cite{Chung} which is built out of
relative momentum.
Thus here we give only covariant tensors that correspond to the
pure S-, P-, D-, and F-wave orbital angular
 momenta:
\begin{eqnarray}\label{l0}
\tilde t^{(0)} &=& 1, \\\label{l1}
 \tilde t^{(1)}_\mu &=& \tilde
g_{\mu\nu}(p_a)r^\nu B_1(Q_{abc}) \equiv\tilde r_\mu
B_1(Q_{abc}),\\\label{l2} \tilde t^{(2)}_{\mu\nu} &=& [\tilde
r_\mu\tilde r_\nu -{1\over 3}(\tilde r\cdot\tilde r)\tilde
g_{\mu\nu}(p_a)]B_2(Q_{abc}), \\\label{l3} \tilde
t^{(3)}_{\mu\nu\lambda} &=& [\tilde r_\mu\tilde r_\nu\tilde
r_\lambda -{1\over 5}(\tilde r\cdot\tilde r)(\tilde
g_{\mu\nu}(p_a)\tilde r_\lambda +\tilde g_{\nu\lambda}(p_a)\tilde
r_\mu+\tilde g_{\lambda\mu}(p_a)\tilde r_\nu)]B_3(Q_{abc}),
\\ & &
\cdots  \nonumber
\end{eqnarray}
where $r=p_b-p_c$  is the relative  four momentum of the two
decay products in the parent particle rest frame; $(\tilde r\cdot\tilde r)=
-{\textbf{r}^2}$.
and
\begin{equation}
\tilde g_{\mu\nu}(p_a) = g_{\mu\nu} -
\frac{p_{a\mu}p_{a\nu}}{p_a^2};
\end{equation}
$B_{L_{bc}}(Q_{abc})$ is a Blatt-Weisskopf barrier factor\cite{Chung,Hippel}, here
$Q_{abc}$ is the magnitude of $\bf p_b$ or $\bf
p_c$ in the rest system of $a$,
\begin{equation}
Q_{abc}^2=\frac{(s_{a}+s_{b}-s_{c})^{2}}{4s_{a}}-s_{b}
\end{equation}
with $s_a=E_a^2-{\bf p}^2_a$.

The spin-1 and spin-2 particles wave functions $\phi_\mu(p_a,m)$
and $\phi_{\mu\nu}(p_a,m)$ satisfy the following conditions
\begin{eqnarray}
&& p_a^{\mu} \phi_\mu(p_a,m)=0,
\hspace{1cm}\phi_\mu(p_a,m)\phi^{*\mu}(p_a,m)=-\delta_{mm'}\nonumber\\\label{proj1}
&&
\sum_m\phi_\mu(p_a,m)\phi^{*}_{\nu}(p_a,m)=-g_{\mu\nu}+\frac{p_{a\mu}p_{a\nu}}{
p_a^2}\equiv -\tilde g_{\mu\nu}(p_a),
\end{eqnarray}
and
\begin{eqnarray}
&&p_a^\mu\phi_{\mu\nu}(p_a,m)=0,\hspace{0.8cm}\phi_{\mu\nu}=\phi_{\nu\mu},\hspace{0.8cm}
 g^{\mu\nu}\phi_{\mu\nu}=0,\hspace{0.8cm}
\phi_{\mu\nu}(m)\phi^{*\mu\nu}(m')=\delta_{mm'}\nonumber\\\label{proj2}
&&P^{(2)}_{\mu\nu\mu'\nu'}(p_a) =
\sum_m\phi_{\mu\nu}(p_a,m)\phi^*_{\mu'\nu'}(p_a,m)={1\over 2}
(\tilde g_{\mu\mu'}\tilde g_{\nu\nu'}+\tilde g_{\mu\nu'}\tilde
g_{\nu\mu'})-{1\over 3}\tilde g_{\mu\nu}\tilde g_{\mu'\nu'}.
\end{eqnarray}

Note that for a given decay process $a\to bc$, the total
angular momentum should be conserved, which means
\begin{equation}\label{conservation}
\textbf{J}_a = \textbf{S}_{bc} + \textbf{L}_{bc},
\end{equation}
where
\begin{equation}
\textbf{S}_{bc} = \textbf{S}_b + \textbf{S}_c .
\end{equation}
In addition parity should also be conserved, which means
\begin{equation}
\eta_a = \eta_b\eta_c (-1)^{L_{bc}},
\end{equation}
where $\eta_a$,  $\eta_b$, and $\eta_c$ are the intrinsic parities
of particles $a$, $b$, and $c$, respectively. From this relation,
one knows whether $L_{bc}$ should be even or odd. Then from Eq.
(\ref{conservation}) one can find out how many different
$L_{bc}-S_{bc}$ combinations, which determine the number of
independent couplings. Also note that in the construction of the
covariant tensor amplitude, for $S_{bc}-L_{bc}-J_a$ coupling, if
$S_{bc}+L_{bc}+J_a$ is an odd number, then
$\epsilon_{\mu\nu\lambda\sigma}p_a^\sigma$ with $p_a$ the momentum
of the parent particle is needed; otherwise it is not needed.

\section{Covariant tensor formalism for $\psi$ decay  into
 $\gamma B\bar B$}
The general form of the decay $\psi \rightarrow \gamma X
\rightarrow \gamma p\bar p$ amplitude can be written as follows by
using the polarization four-vectors of the initial and final
states,
\begin{equation}
A^{(s)}=\psi_\mu(m_J) e^*_\nu(m_\gamma)\psi_{\alpha_s}(p_b,S_b;
p_c,S_c) A^{\mu\nu\alpha_s} =\psi_\mu(m_J)
e^*_\nu(m_\gamma)\psi_{\alpha_s}(p_b,S_b;p_c,S_c)\sum_i\Lambda_i
U_i^{\mu\nu\alpha_s}.
\end{equation}
where $\psi_\mu(m_J)$ is the polarization four vector of the
$\psi$ with spin projection of $m_J$; $e_\nu(m_\gamma)$ is the
polarization four vector of the photon with spin projections of
$m_\gamma$; $U_i^{\mu\nu\alpha_s}$ is the $i$-th partial wave
amplitude with coupling strength determined by a complex parameter
$\Lambda_i$. The spin-1 polarization vector $\psi_\mu(m_J)$ for
$\psi$ with four momentum $p_\mu$ satisfies
\begin{equation}
\sum_{m_{J}=1}^3\psi_\mu(m_J)\psi^{*}_{\nu}(m_J)=
-g_{\mu\nu}+{p_\mu p_\nu\over p^2}\equiv -\tilde g_{\mu\nu}(p),
\end{equation}
with $p^\mu\psi_\mu  = 0$. Here the Minkowsky metric tensor has
the form
$$
g_{\mu\nu} = diag(1,-1,-1,-1).
$$
For $\psi$ production from $e^+ e^-$ annihilation, the electrons
are highly relativistic, with the result that $J_z = \pm 1$ for
the $\psi$ spin projection taking the beam direction as the
z-axis. This limits $m_J$ to 1 and 2, i.e. components along $x$
and $y$. Then one has the following relation
\begin{equation}\label{jpsi}
\sum^2_{m_J=1}\psi_{\mu}(m_J)\psi^*_{\mu'}(m_J)
=\delta_{\mu\mu'}(\delta_{\mu 1}+\delta_{\mu 2}).
\end{equation}
For the  photon polarization four vector, there is the usual
Lorentz orthogonality conditions. Namely, the polarization four
vector $e_\nu(m_\gamma)$ of the  photon with  momenta $q$
satisfies
\begin{equation}
q^\nu e_\nu(m_\gamma)=0,
\end{equation}
which states that spin-1 wave function is orthogonal to its own
momentum.
 The above relation is the same as for a massive vector meson. However, for
the photon, there is an additional gauge invariance condition.
Here we assume the Coulomb gauge in the $\psi$ rest system, {\sl
i.e.}, $p^\nu e_\nu =0$.  Then we have \cite{Greiner}
\begin{equation}\label{photon}
\sum_{m_\gamma} e^*_\mu(m_\gamma) e_\nu(m_\gamma) =
-g_{\mu\nu}+\frac{q_\mu K_\nu+ K_\mu q_\nu}{q\!\cdot\!
K}-\frac{K\!\cdot\! K}{(q\!\cdot\! K)^2}q_\mu q_\nu \equiv
-g^{(\perp\perp)}_{\mu\nu}
\end{equation}
with $K=p-q$ and $K^\nu e_\nu =0$.
 For $X \to p\bar p $,
the total spin of $p\bar p$ system can be either $0$ or $1$. These
two states can be represented by $\psi$ and $\psi_\alpha$
\cite{zou1}. where
\begin{eqnarray}
 \psi &= & \bar u(p_b,S_b)\gamma_5
v(p_c,S_c), \hspace{4cm} if \hspace{0.3cm} s=0 ,\\
\psi_{\alpha} &=&\bar u(p_b,S_b)(\gamma_\alpha -
\frac{r_\alpha}{m_X + 2m} )v(p_c,S_c), \hspace{1.2cm} if
\hspace{0.5cm} s=1 .
\end{eqnarray}
One can see that both  $\psi$ and $\psi_\alpha$ have no dependence
on the direction of the momentum $\hat{\textbf{p}}$, hence
correspond to pure spin states with the total spin of $0$ and $1$,
respectively. Where $p_b$, $p_c$, and $S_b$, $S_c$ are momenta and
spin of the proton antiproton pairs, respectively. $m_X$ and $m$
are the masses of $X$ and  $p$, $\bar p$, respectively;
$u(p_b,S_b)$ and $v(p_c,S_c)$ are the standard Dirac spinor. If we
sum over the polarization, we have the two projection operators:
\begin{eqnarray}\label{diracpo}
\sum_{S_b}u_\alpha(p_b,S_b)\bar u_\beta(p_b,S_b)&=&
\Big(\;\frac{\not\!p_b +
m }{2m}\Big)_{\alpha\beta}\nonumber\\
\sum_{S_c}v_\alpha(p_c,S_c)\bar v_\beta(p_c,S_c)&=&
\Big(\;\frac{\not\!p_c - m }{2m}\Big)_{\alpha\beta}
\end{eqnarray}

To compute the differential cross section, we need an expression
for $|A|^2$. Note that the square modulus of the decay amplitude,
which gives the decay probability  of a certain configuration
should be independent of any particular frame, that is, a Lorentz
scalar. Thus by using the Eqs. (\ref{jpsi}) and (\ref{photon}),
the differential cross section for the radiative decay to an
3-body final state is:
\begin{eqnarray}
\frac{d\sigma^{(s)}}{d\Phi_3}\!&=&\!\frac{1}{2}\sum_{S_b,S_c}\sum^2_{m_J=1}\sum^2_{m_\gamma=1}
|\psi_{\mu}(m_J)
e^*_{\nu}(m_\gamma)\psi_{\alpha_s}(p_b,S_b;p_c,S_c)A^{\mu\nu\alpha_s}|^2
\nonumber\\
&=& -\frac{1}{2}\sum_{S_b,S_c}\sum^2_{\mu=1}A^{\mu\nu\alpha_s}
g^{(\perp\perp)}_{\nu\nu'}A^{*\mu\nu'\alpha'_s}
\psi^{*}_{\alpha_s}\psi_{\alpha'_s}\nonumber\\
&=& -\frac{1}{2}\sum_{i,j}\Lambda_i\Lambda_j^*\sum^2_{\mu=1}
U_i^{\mu\nu\alpha_s}g^{(\perp\perp)}_{\nu\nu'}U_j^{*\mu\nu'\alpha'_s}
\sum_{S_b,S_c}\psi^{*}_{\alpha_s}\psi_{\alpha'_s}\nonumber\\
 &\equiv& \sum_{i,j}P_{ij}\cdot F^{(s)}_{ij}
\end{eqnarray}
where
\begin{eqnarray}
P_{ij} &= P^*_{ji} &= \Lambda_i\Lambda^*_j, \nonumber\\
F^{(s)}_{ij} &= F^{*(s)}_{ji} &= -\frac{1}{2}\sum^2_{\mu=1}
U_i^{\mu\nu\alpha_s}g^{(\perp\perp)}_{\nu\nu'}U_j^{*\mu\nu'\alpha'_s}
\sum_{S_b,S_c}\psi^{*}_{\alpha_s}\psi_{\alpha'_s}.
\end{eqnarray}
$d\Phi_3$ is the standard Lorentz invariant 3-body phase space
given by
\begin{equation}
d\Phi_3(p; q, p_b, p_c)=\delta^4(p - q - p_b - p_c) \frac{d^3{\bf
q}}{(2\pi)^3 2E_\gamma }  \frac{m^2 d^3{\bf p}_b d^3{\bf p}_c }
{(2\pi)^3 E_b (2\pi)^3 E_c }.
\end{equation}
\begin{eqnarray}
F^{(0)}_{ij} &=& F^{*(0)}_{ji} = -\frac{1}{2}\sum^2_{\mu=1}
U_i^{\mu\nu}g^{(\perp\perp)}_{\nu\nu'}U_j^{*\mu\nu'}
\sum_{S_b,S_c}\psi^{*}\psi\nonumber\\
&=&\frac{1}{2}\sum^2_{\mu=1}
U_i^{\mu\nu}g^{(\perp\perp)}_{\nu\nu'}U_j^{*\mu\nu'}
Tr\Big(\;\frac{\not\!p_b + m }{2m}\;\gamma_5 \;\frac{\not\!p_c - m
}{2m}\;\gamma_5 \;\Big)\nonumber\\
&=&-\frac{m^{2}_X}{4m^2}\sum^2_{\mu=1}
U_i^{\mu\nu}g^{(\perp\perp)}_{\nu\nu'}U_j^{*\mu\nu'}.
\end{eqnarray}
The spin sums can be performed using the completeness relations
from Eq.  (\ref{diracpo}):
\begin{eqnarray}
F^{(1)}_{ij} &= &F^{*(1)}_{ji} = -\frac{1}{2}\sum^2_{\mu=1}
U_i^{\mu\nu\alpha}g^{(\perp\perp)}_{\nu\nu'}U_j^{*\mu\nu'\alpha'}
\sum_{S_b,S_c}\psi^{*}_\alpha\psi_{\alpha'}\nonumber\\
&=& -\frac{1}{2} \sum^2_{\mu=1}
U_i^{\mu\nu\alpha}g^{(\perp\perp)}_{\nu\nu'}U_j^{*\mu\nu'\alpha'}
 \Big[ Tr\Big(\;\frac{\not\!p_b + m }{2m}\;\gamma_\alpha
\;\frac{\not\!p_c - m }{2m}\;\gamma_{\alpha'}\;\Big)\nonumber\\
&-& \frac{r_\alpha}{m_X + 2m} Tr\Big(\;\frac{\not\!p_b + m
}{2m}\;\frac{\not\!p_c - m }{2m}\;\gamma_{\alpha'}\;\Big) -
\frac{r_{\alpha'}}{m_X + 2m} Tr\Big(\;\frac{\not\!p_b + m
}{2m}\;\gamma_\alpha \;\frac{\not\!p_c - m }{2m}\;\Big)\nonumber\\
& + & \frac{r_\alpha r_{\alpha'}}{(m_X  + 2m)^2}
Tr\Big(\;\frac{\not\!p_b + m }{2m}\;\frac{\not\!p_c - m
}{2m}\;\Big) \Big]\nonumber\\
&=& -\frac{1}{4m^2} \sum^2_{\mu=1}
U_i^{\mu\nu\alpha}g^{(\perp\perp)}_{\nu\nu'}U_j^{*\mu\nu'\alpha'}
 \Big[p_{b\alpha} p_{b\alpha'}+p_{c\alpha} p_{c\alpha'}+
 p_{b\alpha} p_{c\alpha'} +  p_{b\alpha'} p_{c\alpha}
 - m^{2}_X g_{\alpha\alpha'}
 \Big].\nonumber\\
\end{eqnarray}
where we have used traces of $\gamma$ matrics,
\begin{eqnarray}
&& tr(1) = 4, \hspace{1.2cm} tr(\gamma^5) = 0, \hspace{1.5cm}
tr(\gamma^\mu \gamma^\nu) = 4 g^{\mu\nu} \nonumber\\
&& tr( any \;\;odd \;\; \sharp\;\; \gamma's) = 0, \hspace{2cm}
tr(\gamma^\mu\gamma^\nu\gamma^5) = 0.
\end{eqnarray}

\subsection*{Amplitudes for the radiative decay
$\psi  \rightarrow \gamma p\bar p$}

We consider the decay of a $\psi$ state into two steps:
$\psi\to\gamma X$ with $X\to p\bar p$. The possible $J^{PC}$ for
$X$ are $0^{++}, 0^{-+}$, $1^{++}$, $2^{++}$, $2^{-+}$, etc. For
$\psi\to\gamma X$, we choose two independent momenta $p$ for
$\psi$ and $q$ for the photon to be contracted with spin wave
functions. We denote the four momentum of $X$  by $K$. The tensor
describing the first and second steps will be denoted by $\tilde
T^{(L)}_{\mu_1\cdots\mu_L}$ and $\tilde
t^{(\textit{l})}_{\mu_1\cdots\mu_{\textit{l}}}$, respectively.

For $\psi\to\gamma 0^{++} \rightarrow \gamma p\bar p$, there is
one independent covariant tensor amplitude:
\begin{equation}\label{0p}
U^{\mu\nu\alpha} = g^{\mu\nu}  \tilde t^{(1)\alpha}.
\end{equation}

For $\psi\to\gamma 0^{-+} \rightarrow \gamma p\bar p$, there is
one independent covariant tensor amplitude:
\begin{equation}\label{0m}
U^{\mu\nu}=  \epsilon^{\mu\nu\lambda\sigma}p_\lambda q_\sigma
B_1(Q_{\psi\gamma X}).
\end{equation}

 For $\psi\to\gamma 1^{++} \rightarrow
\gamma p\bar p$, there are two independent covariant tensor
amplitudes:
\begin{eqnarray}\label{1p1}
U^{\mu\nu\alpha}_1 \!&=&\! \epsilon^{\mu\nu\lambda\sigma}p_\lambda
\epsilon^{\alpha\beta\rho}_{\;\;\;\;\;\;\sigma}K_\rho \tilde
t^{(1)}_\beta  ,\\\label{1p2}U^{\mu\nu\alpha}_ 2 \!&=&\!
\epsilon^{\nu\lambda\sigma\gamma}p_\lambda q^{\mu} q_{\gamma}
\epsilon^{\alpha\beta\rho}_{\;\;\;\;\;\;\sigma}K_\rho \tilde
t^{(1)}_\beta B_{2}(Q_{\psi\gamma X}).
\end{eqnarray}

For $\psi\to\gamma 1^{-+}$, the exotic $1^{-+}$ meson cannot decay
into $p\bar p$.

For $\psi\to\gamma 2^{++} \rightarrow  p\bar p$, there are three
independent covariant tensor amplitudes:
\begin{eqnarray}\label{2p1}
U^{\mu\nu\alpha}_1\!&=&\! P^{(2)\mu\nu\alpha\beta}(K)\tilde t^{(1)}_\beta
,\\\label{2p2} U^{\mu\nu\alpha}_2\!&=&\!
 g^{\mu\nu} P^{(2)\alpha\beta\lambda\sigma}
p_{\beta}p_{\lambda}
 \tilde t^{(1)}_\sigma B_{2}(Q_{\psi\gamma X})
,\\\label{2p3}
 U^{\mu\nu\alpha}_3\!&=&\! P^{(2)\nu\alpha\beta\lambda}
 q^{\mu}p_{\beta}
\tilde t^{(1)}_\lambda B_{2}(Q_{\psi\gamma X}).
\end{eqnarray}
For $2^{++}$ decaying to $p\bar p$, the orbital angular momentum
between the proton and antiproton  $\textit{l}$ could be $1$ and
$3$; but we ignore $\textit{l} =3$ contribution because of the
strong centrifugal barrier.

For $\psi\to\gamma 2^{-+} \rightarrow  p\bar p$,  the possible
partial wave amplitudes are the following:
\begin{eqnarray}\label{2m1}
U^{\mu\nu}_ 1 \!&=&\! \epsilon^{\mu\nu\lambda\sigma}p_\lambda
q^{\gamma} \tilde t^{(2)}_{ \gamma\sigma} B_{1}(Q_{\psi\gamma X}),
 \\\label{2m2}
U^{\mu\nu}_ 2 \!&=&\! \epsilon^{\mu\nu\lambda\sigma}p_\lambda
q_{\sigma}p_{\gamma}p_{\delta} \tilde t^{(2)\gamma\delta}
B_{3}(Q_{\psi\gamma X}),
 \\\label{2m3}
U^{\mu\nu}_ 3 \!&=&\! \epsilon^{\nu\gamma\lambda\sigma}p_\lambda
q_{\sigma}q^{\mu}p^{\delta} \tilde t^{(2)}_{\gamma\delta}
B_{3}(Q_{\psi\gamma X}).
\end{eqnarray}

It is worth to mention here that the above partial wave amplitudes
for the process $J/\psi \to \gamma p\bar p$ are applicable to the
processes $J/\psi \to \gamma\Lambda\bar \Lambda, \gamma\Sigma
\bar\Sigma$, and $\gamma\Xi\bar \Xi$ as well.

\section{Monte Carlo simulation for $J/\psi \to \gamma p\bar p$}

We perform Monte Carlo simulation for the decay process $J/\psi
\to \gamma X \to \gamma p\bar p$, with the $J^{PC}$ of $X$ to be
$0^{++}$, $0^{-+}$, $1^{++}$, $2^{++}$, $2^{-+}$. The predicted
angular distributions for various $J^{PC}$ intermediate resonances
could serve as a useful reference for people performing partial
wave analysis of $\psi\to\gamma B\bar B$ channels.

\begin{figure}\label{0pp}
\centerline{\epsfysize 3 true in \epsfbox{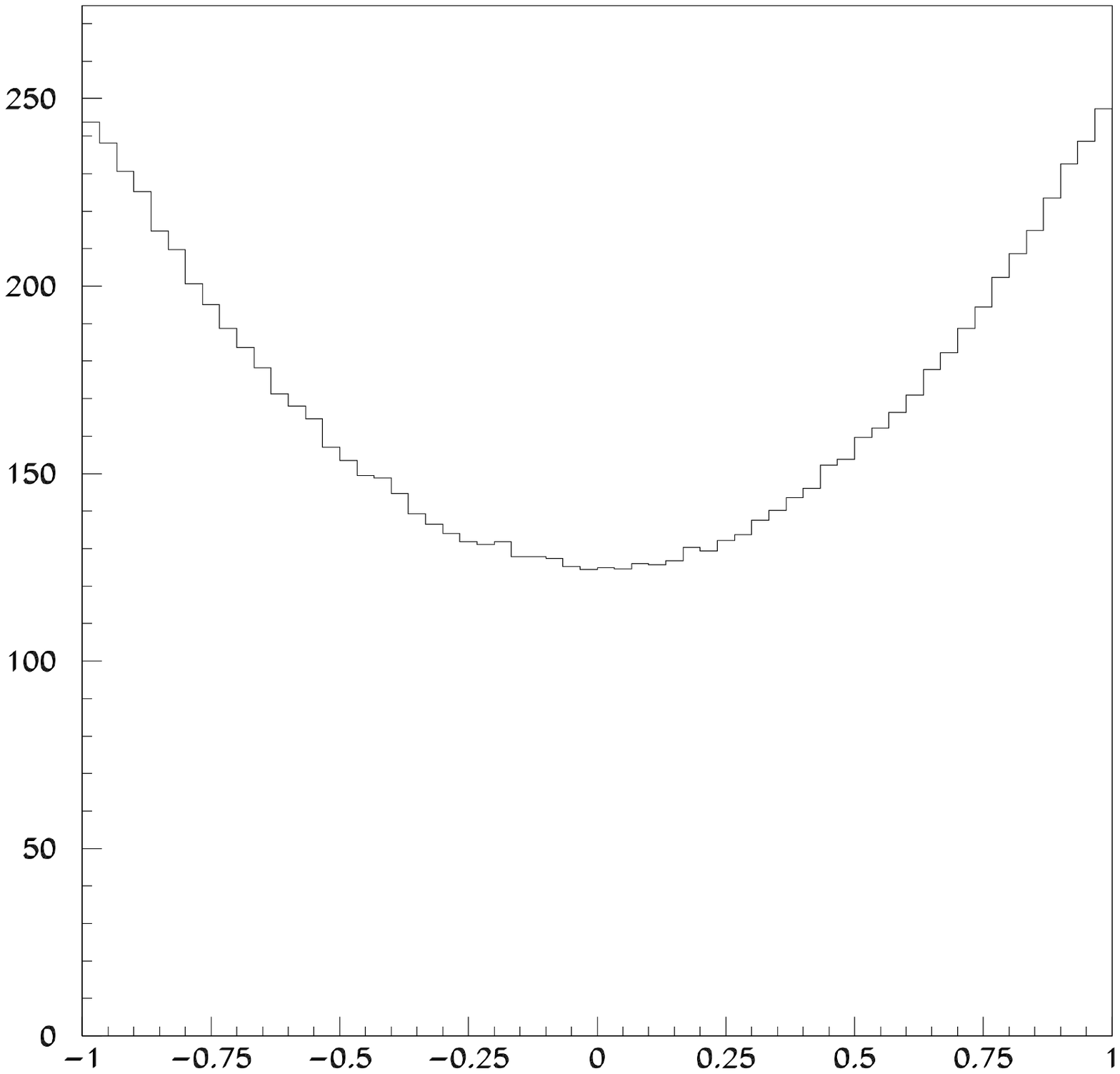}\hspace{2cm}
            \epsfysize 3 true in \epsfbox{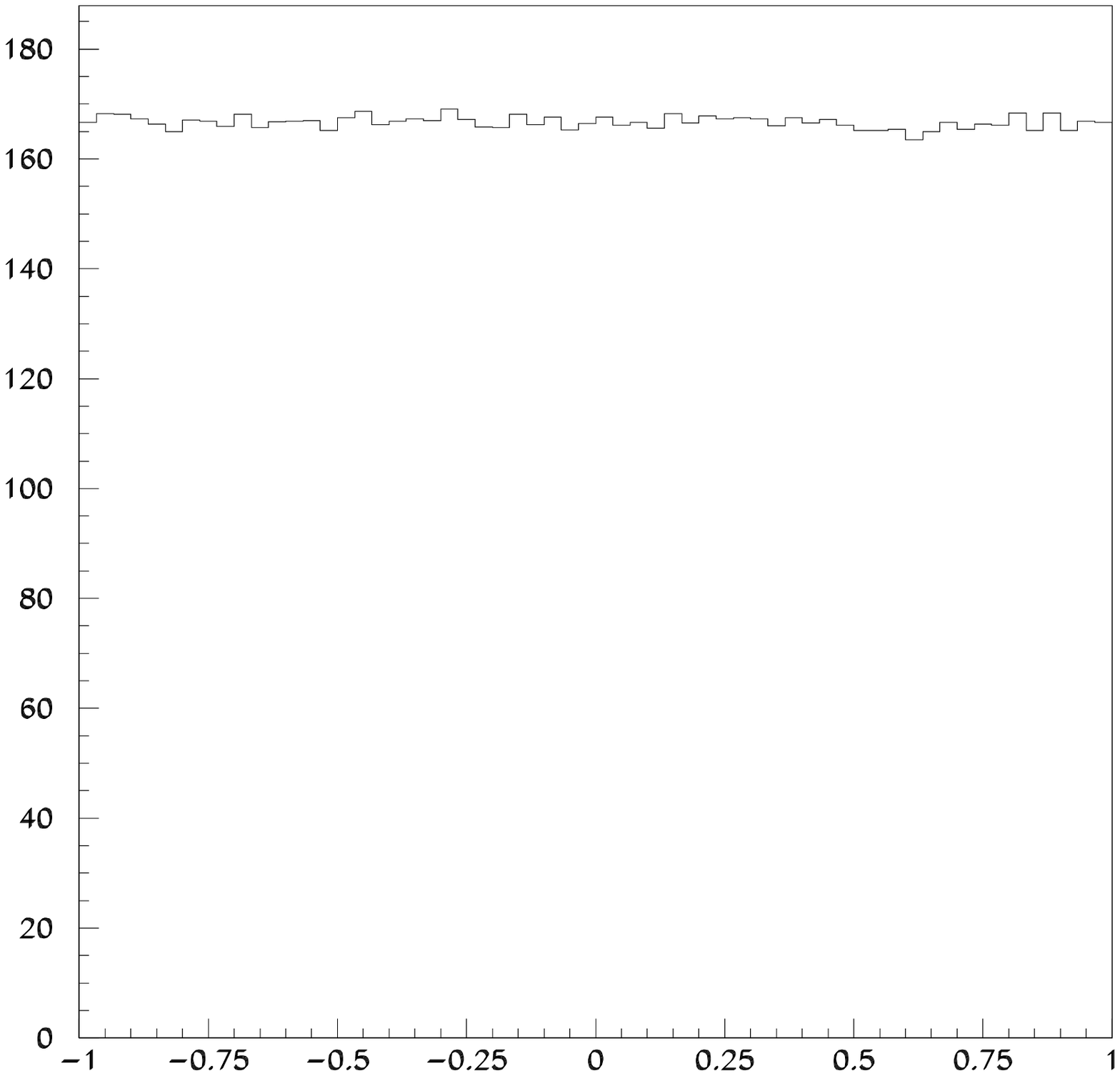}}
\caption{Angular distributions of the photon and proton in the
process $J/\psi \to \gamma 0^{\pm+} \to \gamma p\bar p$.}

\centerline{\epsfysize 3 true in
\epsfbox{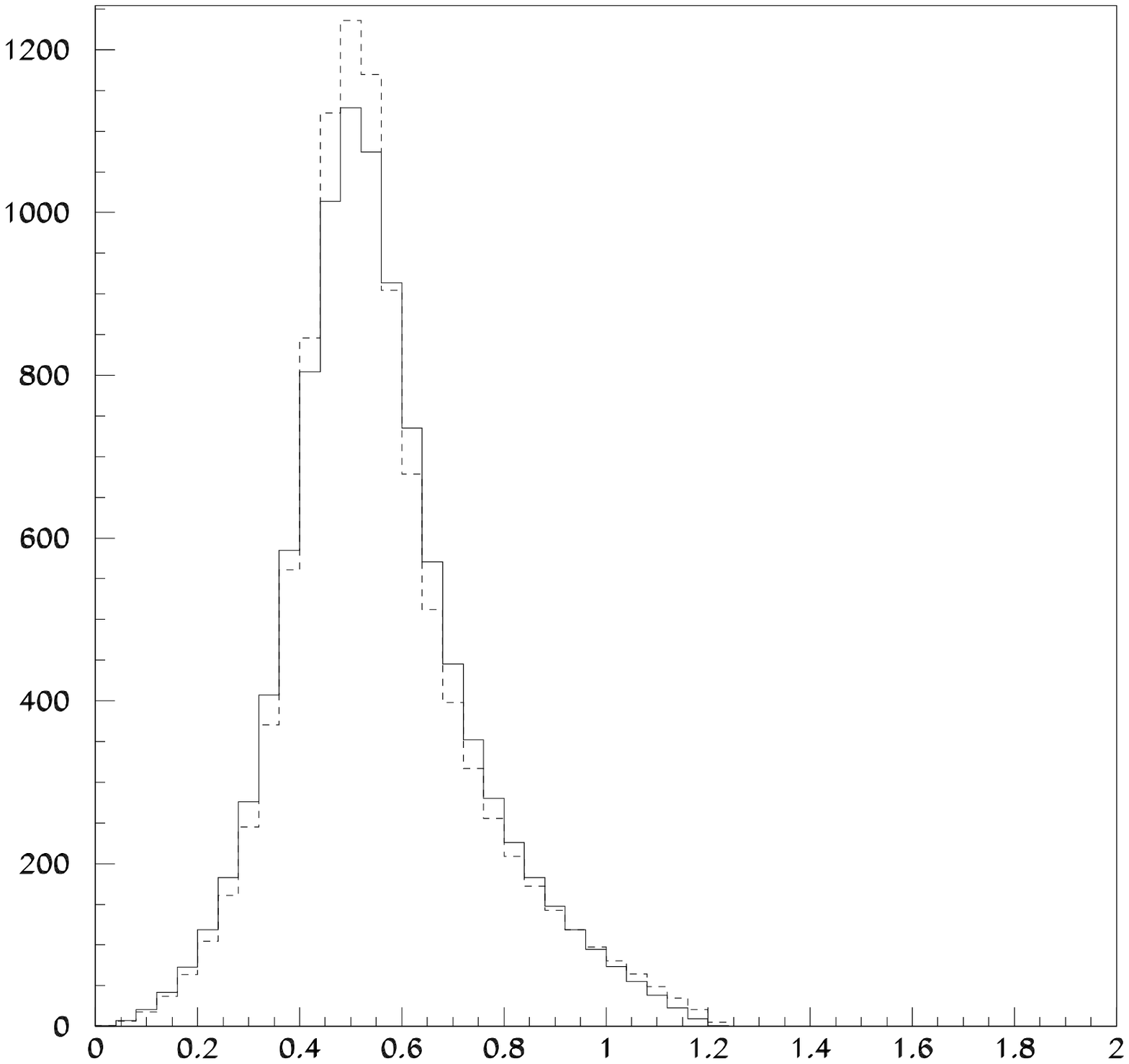}\hspace{2cm}
            \epsfysize 3 true in \epsfbox{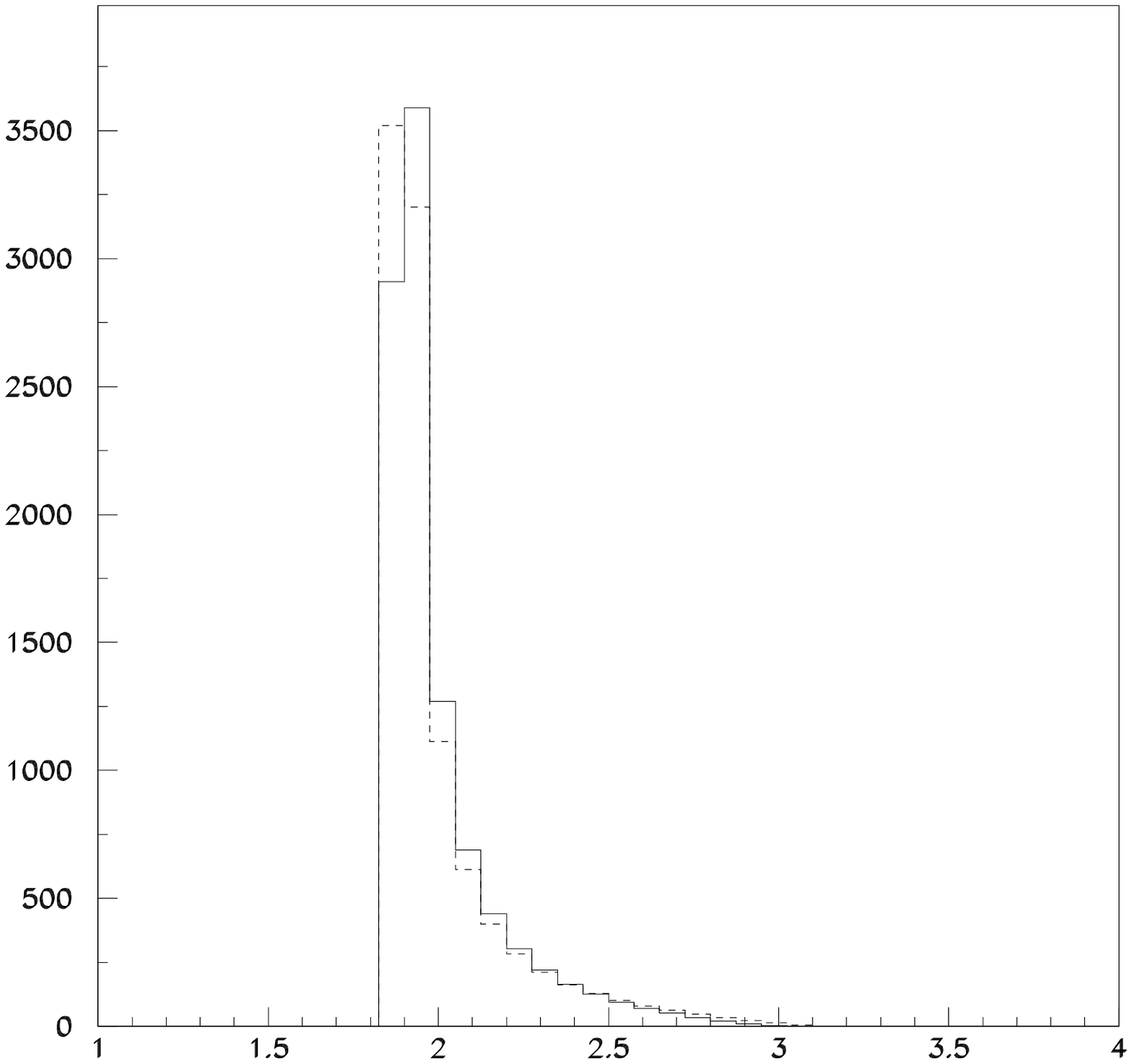}}
 \caption{Momentum distributions of proton and invariant
mass of $p\bar p$ in the process $J/\psi \to \gamma 0^{\pm+} \to
\gamma p\bar p$. The dashed curve is the case of $0^{++}$;the
solid curve is the case of $0^{-+}$.}
 \begin{picture}(800,140)
 \put(105,470){cos$\theta_{\gamma}$}
 \put(325,470){cos$\theta_{p}$}
 \put(100,210){P$_{p}(GeV)$}
 \put(310,210){M$_{p\bar p}(GeV)$}
 \end{picture}
\end{figure}

BES2 reported observation of a narrow enhancement near $2m_p$ in
the invariant mass spectrum of $p\bar p$ pairs from radiative
$J/\psi \to \gamma p\bar p$ decays. The peak has properties
consistent  with either a $J^{PC} = 0^{-+}$ or $0^{++}$. We
simulate these two processes with the Breit-Wigner mass and width
of $X$ in Ref.\cite{BES}. The angular distributions of the photon
and proton are shown in Fig.1, while the invariant mass of $p\bar
p$ and momentum distribution of the proton are shown in Fig.2. The
two fits of $0^{-+}$ and $0^{++}$ are indeed hardly
distinguishable from these distributions. However, if the narrow
structure is really due to a narrow $0^{-+}$ or $0^{++}$
resonance, it can be distinguished by its other decay modes. While
a $0^{-+}$ resonance can decay into $\eta\pi\pi$, $K\bar K\pi$, a
$0^{++}$ resonance cannot. On the other hand, while a $0^{++}$
resonance can decay into two pseudoscalar mesons, such as
$\pi\pi$, $K\bar K$, a $0^{-+}$ resonance cannot. Both $0^{-+}$
and $0^{++}$ resonances can decay into $4\pi$ and $\pi\pi K\bar K$
channels. Previous data on these channels
\cite{BES1,BES2,BES3,BES4,BES5,BES6} have not seen such narrow
structure around $2m_p$. This gives support to the explanation of
$p\bar p$ $0^{-+}$ final state interaction \cite{zouchiang,FSI}
for the observed narrow peak structure in $p\bar p$ channel only.

The processes with other $J^{PC}$ intermediate X resonances are
simulated by assuming the mass of X at 2.15 GeV with a width of
0.15 GeV. The angular distribution of photon and proton are shown
in Figs.3-5. The $\theta_\gamma$ and $\theta_p$ are given in
$J/\psi$ rest frame and $X$ rest frame, respectively. The
$\theta_\gamma$ angular distributions coincide with general
analytic formulae given in Ref.\cite{zou2} as it should be. One
can see that various partial wave amplitudes give different
angular distributions. By fitting the theoretical differential
cross section given by Eq.(20) with parameters $\Lambda_i$ to the
data, one can get the magnitudes of each partial wave
contribution.

\begin{figure}\label{1pp}
\centerline{\epsfysize 3 true in \epsfbox{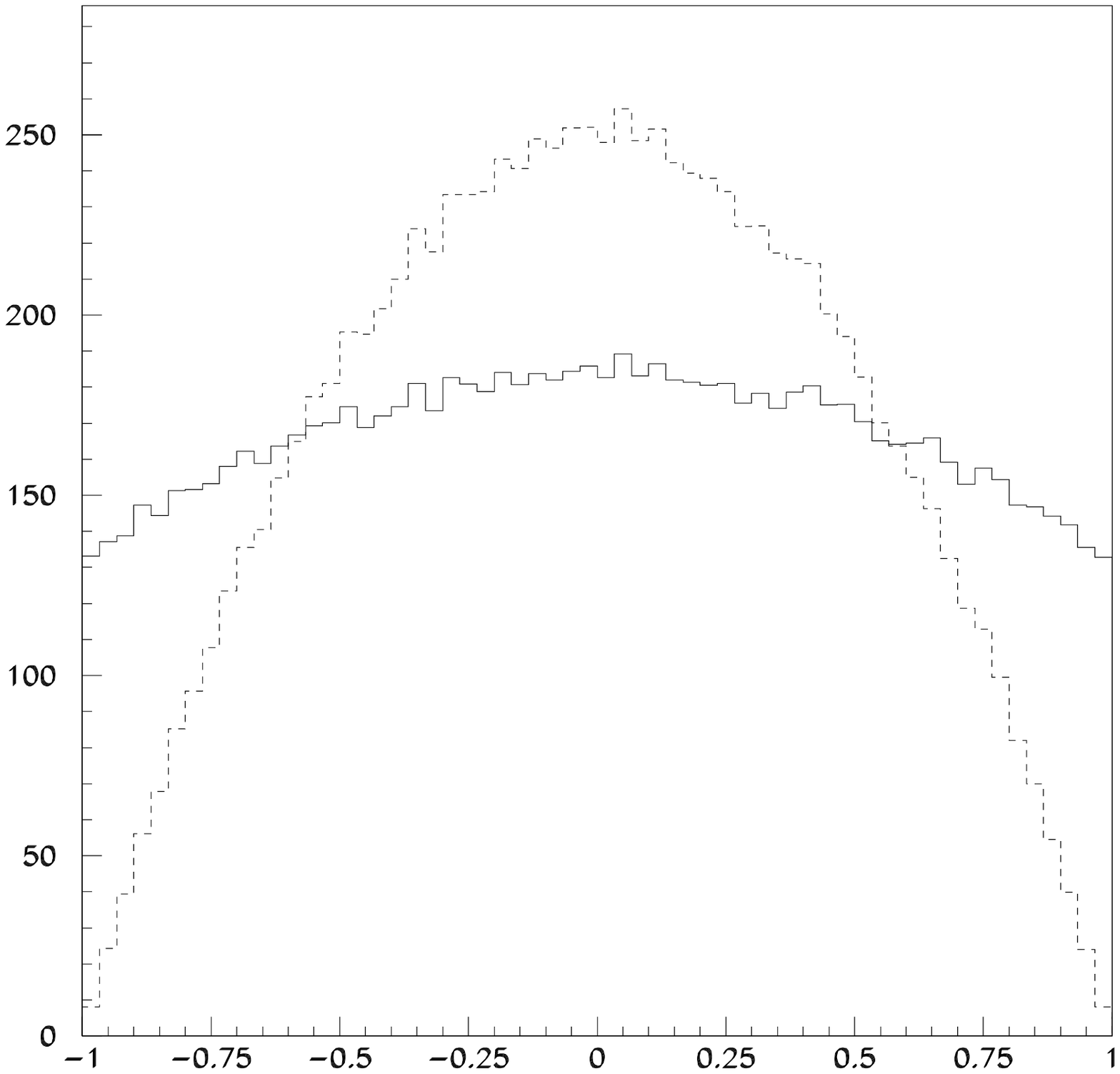}\hspace{2cm}
            \epsfysize 3 true in \epsfbox{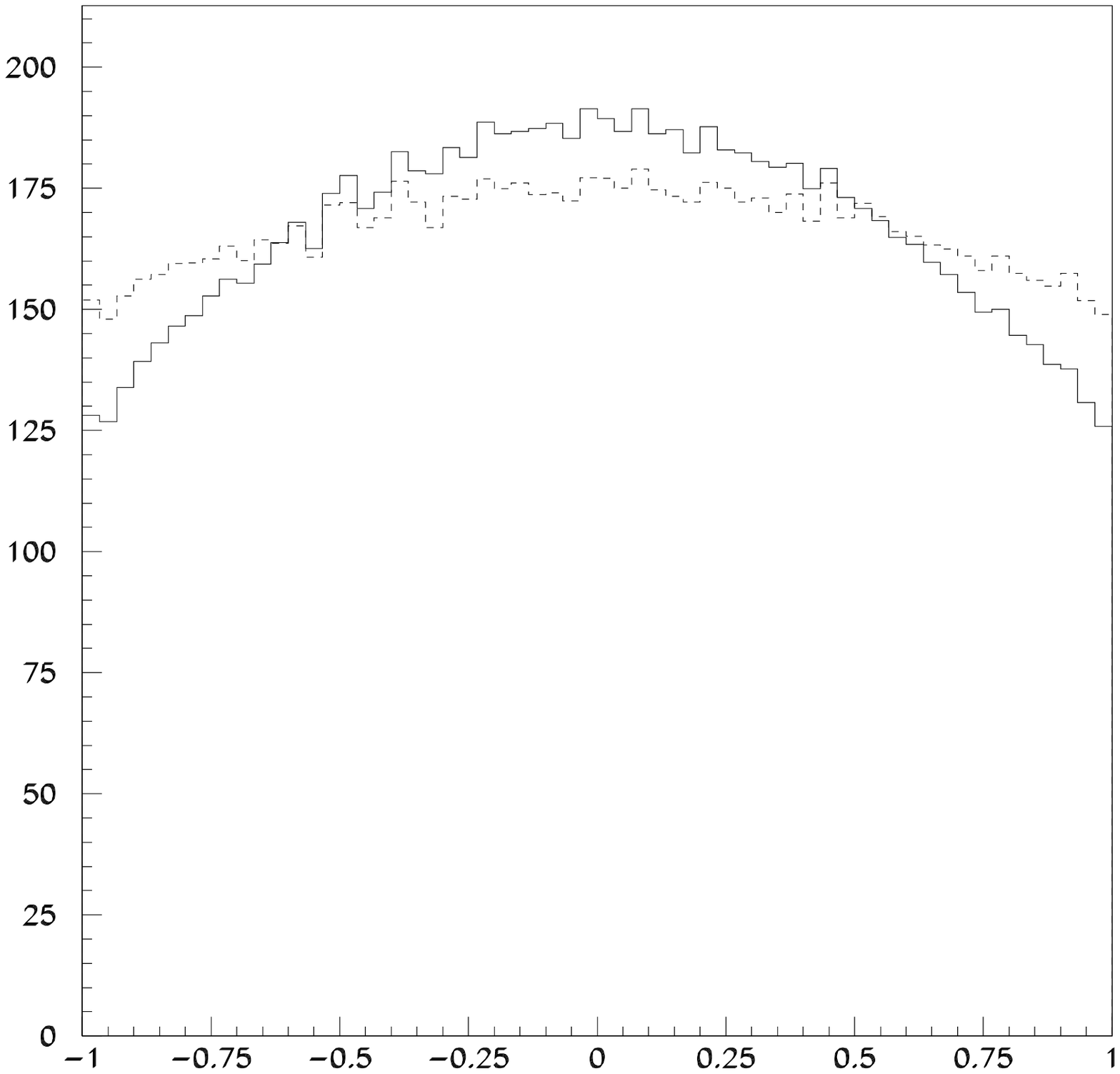}}
\caption{Angular distributions of the photon and proton for
independent amplitudes of the process $J/\psi \to \gamma 1^{++}
\to \gamma p\bar p$. The solid curve and dashed curve correspond
to amplitudes (\ref{1p1}) and (\ref{1p2}), respectively.}
\label{2pp} \centerline{\epsfysize 3 true in
\epsfbox{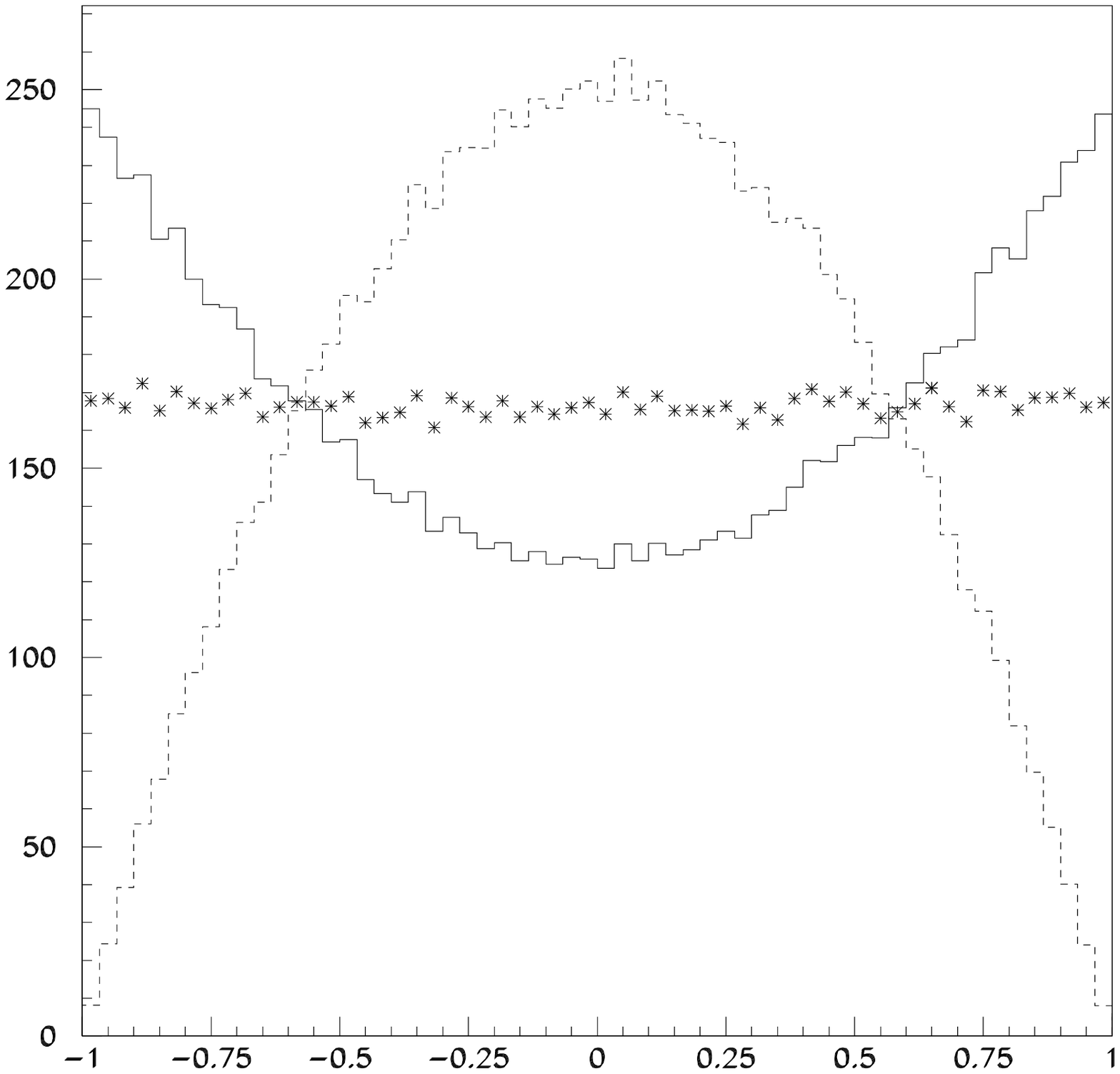}\hspace{2cm}
            \epsfysize 3 true in \epsfbox{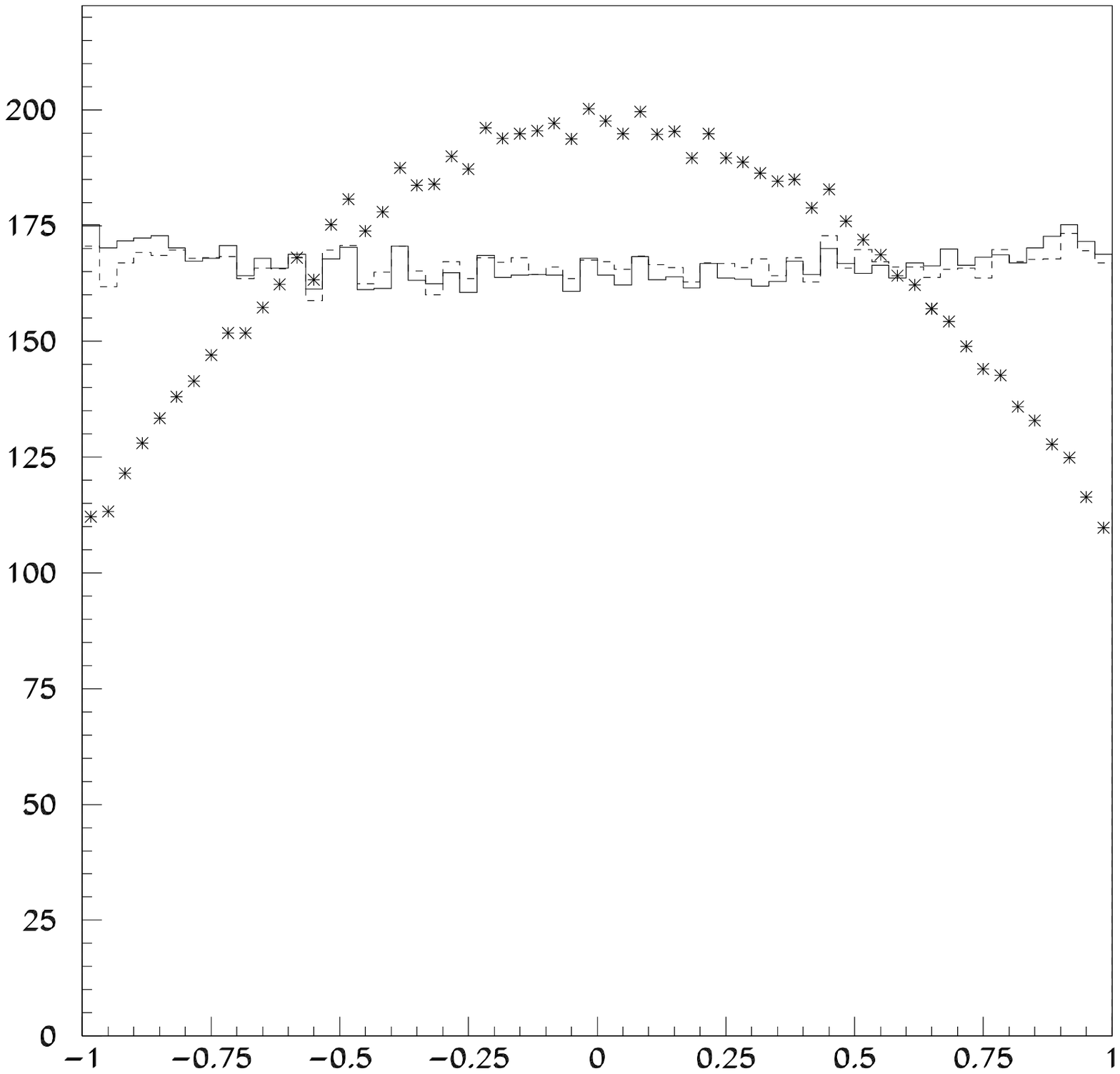}}
\caption{Angular distributions of the photon and proton given by
independent amplitudes Eq.(\ref{2p1}) (stars), Eq.(\ref{2p2})
(solid lines),and Eq.(\ref{2p3}) (dashed lines) of the process
$J/\psi \to \gamma 2^{++} \to \gamma p\bar p$.}
\begin{picture}(800,140)
 \put(105,215){cos$\theta_{\gamma}$}
 \put(325,215){cos$\theta_{p}$}
 \put(105,490){cos$\theta_{\gamma}$}
 \put(325,490){cos$\theta_{p}$}
\end{picture}
\end{figure}
\begin{figure}\label{3mp}
\centerline{\epsfysize 3 true in \epsfbox{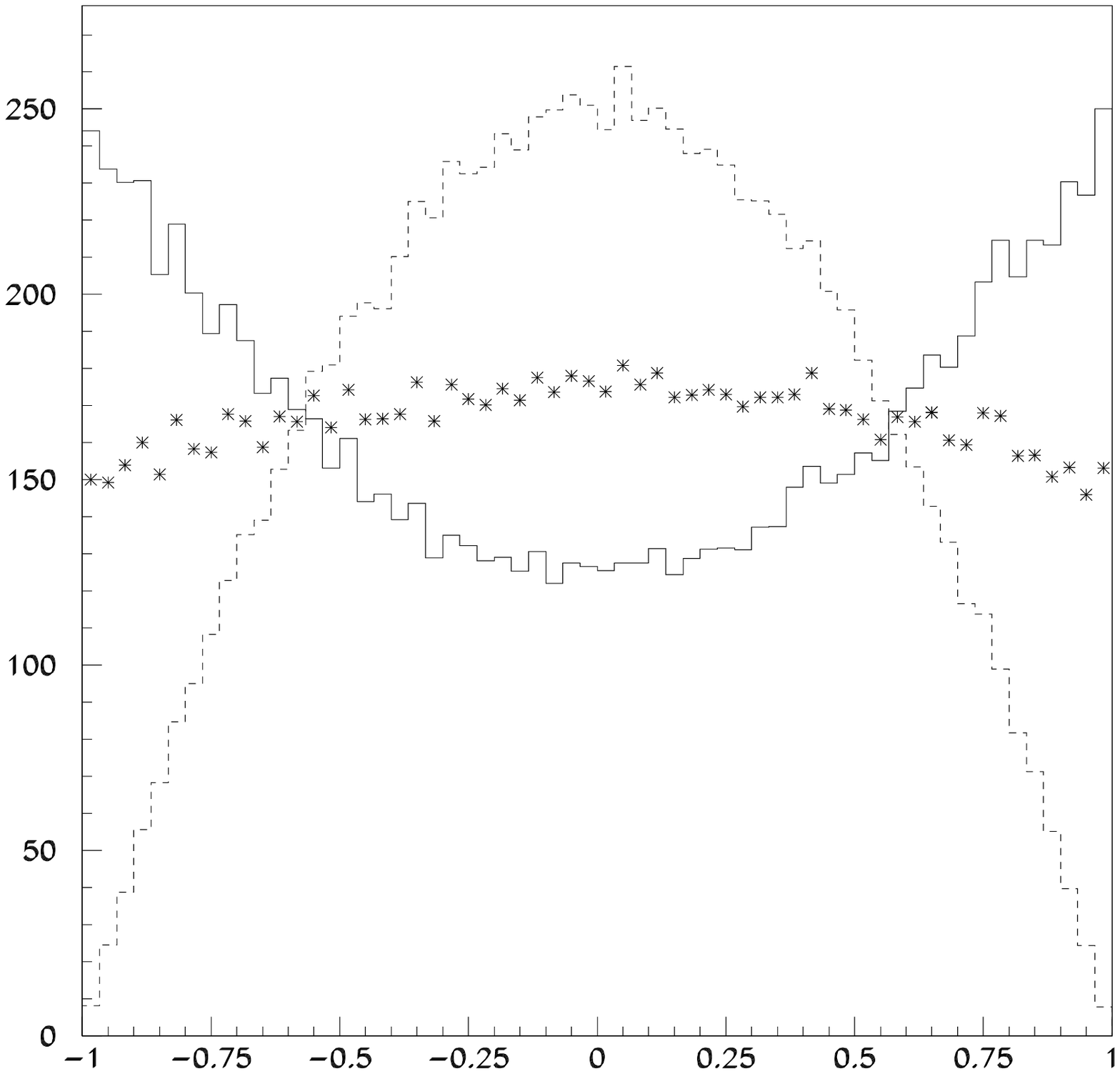}\hspace{2cm}
\epsfysize 3 true in \epsfbox{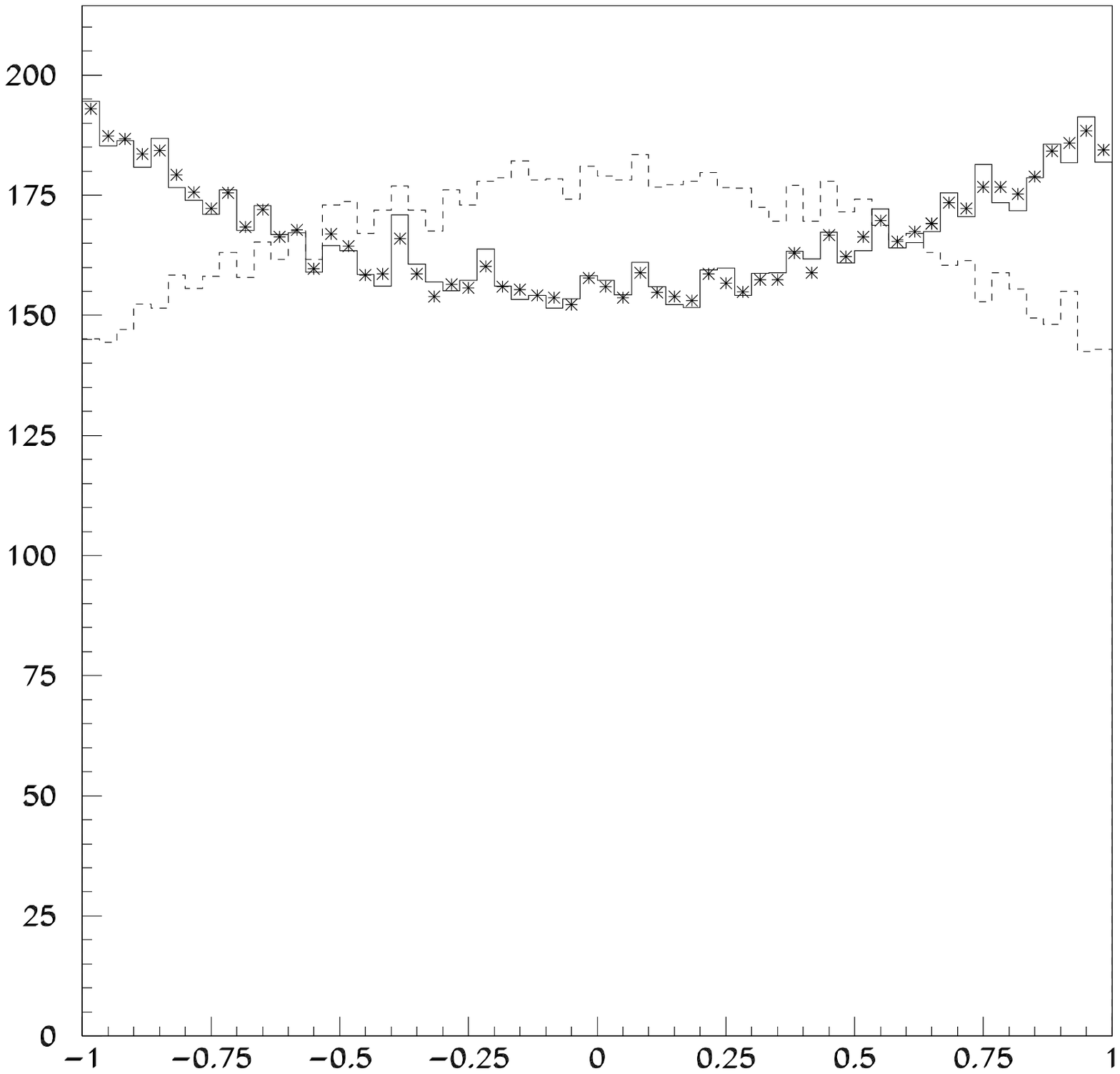}} \caption{Angular
distributions of the photon and proton given by independent
amplitudes Eq.(\ref{2m1}) (stars), Eq.(\ref{2m2}) (solid
lines),and Eq.(\ref{2m3}) (dashed lines) of the process $J/\psi
\to \gamma 2^{-+} \to \gamma p\bar p$.}
\begin{picture}(400,540)
\put(107,615){cos$\theta_{\gamma}$} \put(325,615){cos$\theta_{p}$}
\end{picture}
\end{figure}
\section{Conclusion}

In this paper we provided a theoretical formalism and a Monte
Carlo study of the partial wave analysis for the radiative decay
$J/\psi \to \gamma p \bar p$, which are also applicable to the
processes $J/\psi \to \gamma\Lambda\bar \Lambda, \gamma\Sigma
\bar\Sigma$ and $\gamma\Xi\bar \Xi$. We have constructed all
possible covariant tensor amplitudes for intermediate resonant
states of $J\leq 2$. For intermediate resonant states of $J\geq
3$, the production vertices need $L\geq 2$ and are expected to be
suppressed \cite{zou}. The formulae here can be directly used to
perform partial wave analysis of forthcoming high statistics data
from CLEO-c and BES-III on these channels to extract useful
information on the baryon-antibaryon interactions.

\section*{Acknowledgements}
S. Dulat would like to thank Prof. Kaoru Hagiwara, who has given
productive comments. Both the presentation at KEK and content of
this paper has greatly benefited  from his insightful comments.
The work is partly supported by CAS Knowledge Innovation Project
(KJCX2-SW-N02) and the National Natural Science Foundation of
China under Grant Nos.10225525,10055003, 90103012, 10265003.

\end{document}